# Quasi-epitaxial growth of BaTiS$_3$ films

Mythili Surendran[1,2], Boyang Zhao[1], Guodong Ren[3], Shantanu Singh[1], Amir Avishai[2], Huandong Chen[1], Jae-Kyung Han[4], Megumi Kawasaki[4], Rohan Mishra[3,5] and Jayakanth Ravichandran[1,2,6, a]

[1]Mork Family Department of Chemical Engineering and Materials Science, University of Southern California, 925 Bloom Walk, Los Angeles, CA 90089, USA

[2]Core Center of Excellence in Nano Imaging, University of Southern California, 1002 West Childs Way, MCB Building, Los Angeles, CA 90089, USA

[3]Institute of Materials Science & Engineering, Washington University in St. Louis, One Brookings Drive, St. Louis, MO 63130, USA

[4]School of Mechanical, Industrial & Manufacturing Engineering, Oregon State University, 306 Dearborn, Corvallis, OR 97331, USA

[5]Department of Mechanical Engineering & Materials Science, Washington University in St. Louis, One Brookings Drive, St. Louis, MO 63130, USA

[6]Ming Hsieh Department of Electrical and Computer Engineering, University of Southern California, 925 Bloom Walk, Los Angeles, CA 90089, USA

[a] E-mail: j.ravichandran@usc.edu

## Abstract

Perovskite chalcogenides have emerged as a new class of semiconductors with tunable band gap in the visible-infrared region. High quality thin films are critical to understand the fundamental properties and realize the potential applications based on these materials. We report growth of quasi-epitaxial thin films of quasi one-dimensional (quasi-1D) hexagonal chalcogenide BaTiS$_3$ by pulsed laser deposition. We identified the optimal growth conditions by varying the growth parameters such as the substrate temperature and H$_2$S partial pressure and examined their effects on the thin film structure. High resolution thin film X-Ray diffraction shows strong texture in the out-of-plane direction, whereas no evidence of in-plane relationship between the film and the substrate is observed. Grazing incidence wide-angle X-ray scattering and scanning transmission electron microscopy studies reveal the presence of weak epitaxial relationships of the film and the substrate, despite a defective interface. Our study opens up a pathway to realize quasi-1D hexagonal chalcogenide thin films and their heterostructures with perovskite chalcogenides.

**Introduction**

Mid-infrared materials are of substantial interest due to their broad range of applications such as molecular fingerprint imaging, optical telecommunication, and remote sensing[1-4]. IR materials with intrinsic optical anisotropy can enable novel photonic devices such as polarizing filters, light modulators, and polarization sensitive photodetectors[5-7]. Several two- dimensional (2D) materials such as black phosphorus[8] have been investigated for their optical anisotropy in the mid-infrared energies. Although the anisotropy of 2D materials between the in-plane and out-of-plane is appreciable, the easily accessible in-plane anisotropy is modest. Despite promising developments, the growth of large area mid-infrared responsive 2D materials such as black phosphorous has been a challenge[9].

Chalcogenide perovskites have been studied extensively in the recent past and have indeed emerged as a new class of semiconductors that demonstrate exciting electronic and optical properties[10-14]. Even at this early stage of research, some of the chalcogenide perovskites show promise for technological applications such as photovoltaics, infrared imaging and detection, and thermoelectrics[14-17]. Chalcogenide perovskites with Zr atom at the B-site (in the $ABX_3$ perovskite structure (where *A* is a larger cation than *B*, and *X* = S, Se, or Te) such as $BaZrS_3$ (BZS), $SrZrS_3$ and related layered phases such as Ruddlesden Popper $Ba_3Zr_2S_7$ are largely studied and possess high absorption coefficient[18-20], long recombination lifetimes[21], and excellent thermal stability[22] rendering them as suitable candidates for solar energy conversion and visible optoelectronic applications. Meanwhile, chalcogenide perovskites with Ti atom at the B-site such as BaTiS$_3$ and Sr$_{1+x}$TiS$_3$ are considered candidates for optical elements such as polarizers due to their intrinsic optical anisotropy in the mid-wave (MWIR) and long-wave infrared (LWIR) region[15, 17, 23]. BaTiS$_3$ (BTS) has a quasi-1D hexagonal structure, where parallel chains formed by face-sharing TiS$_6$ octahedra are arranged in hexagonal symmetry. Single crystals of BTS have been demonstrated to possess giant birefringence[15] in the MWIR and LWIR region due to the large structural and chemical anisotropy between intra- and interchain directions. This large intrinsic in-plane anisotropy makes this an accessible material system for polarization dependent infrared imaging and detection. However, further studies are necessary to understand the fundamental transport properties of BTS such as carrier type, density, and mobility. In addition, BTS is a small band-gap non-degenerate semiconductor which could host exciting novel and emergent phenomena in thin film heterostructures and superlattices[24]. Therefore, development of high-quality epitaxial thin film technology is critical to further investigate the fundamental optical and electrical properties of BTS and realize functional advanced photonic devices.

Synthesis of high-quality chalcogenide perovskite thin films presents various challenges such as chalcogen-cation partial pressure mismatch, slow diffusion kinetics and use of highly corrosive and reactive chalcogen precursors. Despite these challenges, epitaxial BZS films[19, 20] were recently demonstrated by pulsed laser deposition and molecular beam epitaxy and show promising optoelectronic properties ideal for future photovoltaic applications. However, thin film growth of other perovskite chalcogenides still remains unexplored.

In this article, we report the growth of quasi-epitaxial BTS thin films by pulsed laser deposition (PLD) on single crystal oxide substrates. We carried out detailed growth studies by varying the substrate temperature and the Ar-$H_2$S partial pressure to understand the effect of these process parameters on the structure of BTS films and achieve highly textured films of $BaTiS_3$. We performed structural and surface characterization of the as-grown films using high-resolution X-ray diffraction, X-ray reflectivity, atomic force microscopy, and grazing incidence wide angle X-ray scattering and scanning transmission electron microscopy.

## Results and Discussion

BTS thin films were grown from a phase pure stoichiometric BTS target in a background gas mixture of hydrogen sulfide (4.75 mol%) and argon (Ar-$H_2$S) on single crystal oxide substrates. Due to the high propensity for the loss of sulfur at high temperatures, conventional sintering methods may result in the formation of nonstoichiometric chalcogenide perovskites. Therefore, a high-pressure torsion (HPT)[25, 26] method that allows room temperature densification was employed to prepare a dense, polycrystalline BTS target (density > 90%) for PLD growth. We varied temperature and the Ar-$H_2$S partial pressure to determine the optimized growth parameters and grew the films on different substrates to understand compatibility of the oxide substrates for these harsh conditions. First, BTS films were grown on various single crystal oxide substrates such as $SrTiO_3$ (STO) (001), GdScO$_3$ (GSO) (110), LaAlO$_3$ (LAO) (001), LSAT (001) and *c*-plane sapphire, at a fixed temperature of 700°C and an Ar-H$_2$S partial pressure of 20 mTorr.

Structural information obtained from high resolution X-ray diffraction (HRXRD) studies was used as the primary criterion for optimizing the growth parameters for BTS thin films. All the characterization were performed on as-grown films without any further annealing. Strong out-of-plane textured BTS films were observed on perovskite oxide substrates STO and GSO, with the strongest texture on STO (001), while polycrystalline films were observed on *c*-plane sapphire substrates. Films grown on other perovskite oxide substrates such as LAO and LSAT did not show an out-of-plane texture in XRD. Figure 1 shows HRXRD scans of BTS films grown on various substrates. BTS single crystal has a quasi-1D hexagonal crystal structure with lattice parameters, $a = b = 6.756$ Å and $c = 5.798$ Å. It is reported to grow in two different orientations such as *a*-axis out-of-plane with *b-c* in-plane orientation, or *c*-axis out-of-plane with *a-b* in-plane orientation[27]. In our case, we observed BTS 110 out-of-plane texture in the films grown on STO and GSO substrates. As the strongest texture was observed on STO, we examined further the effect of the growth conditions on the microstructure of BTS films grown on STO substrates in detail.

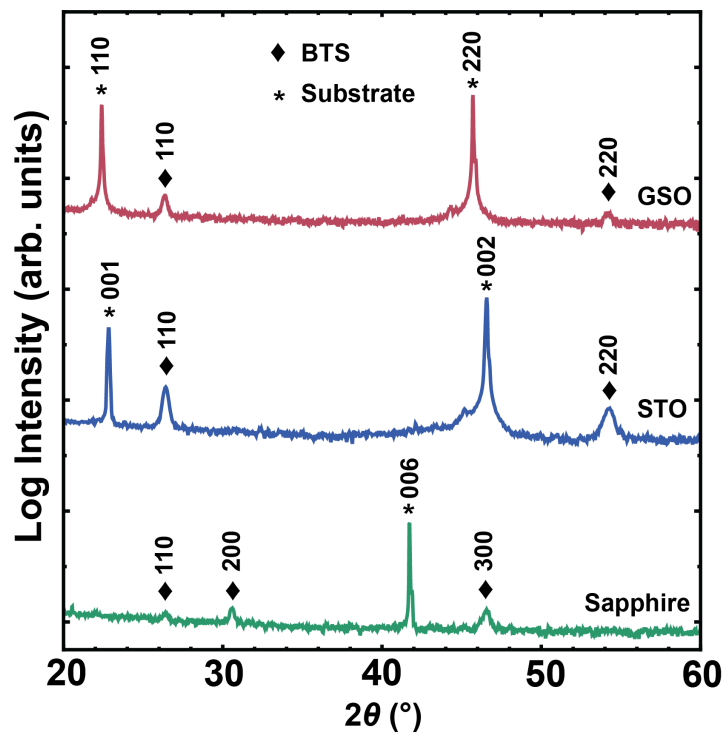

**Figure 1**. High resolution 2$\theta$-$\theta$ XRD patterns of representative BTS films grown on GSO, STO and sapphire substrates.

BTS films were grown at substrate temperatures of 600 – 750°C at a fixed Ar-H$_2$S partial pressure of 20 mTorr. Figure 2a shows HRXRD plot of ~ 100 nm thick films deposited at 600, 650, 700 and 750°C. The BTS films showed strong (110)-oriented texture along the *c*-axis of the STO substrate. At 600°C, the films are likely to be amorphous or poorly textured as we did not observe any clear 110 type reflections in the XRD scans. Films grown above 650°C were crystalline and textured as indicated by the BTS 110 reflection observed at 2$\theta$ of ~26.4°. The strongest texture was observed at 700°C, with the peak intensity decreasing at 750°C and above. BTS films grown at all temperatures were relaxed and only *hh0* reflections were seen.

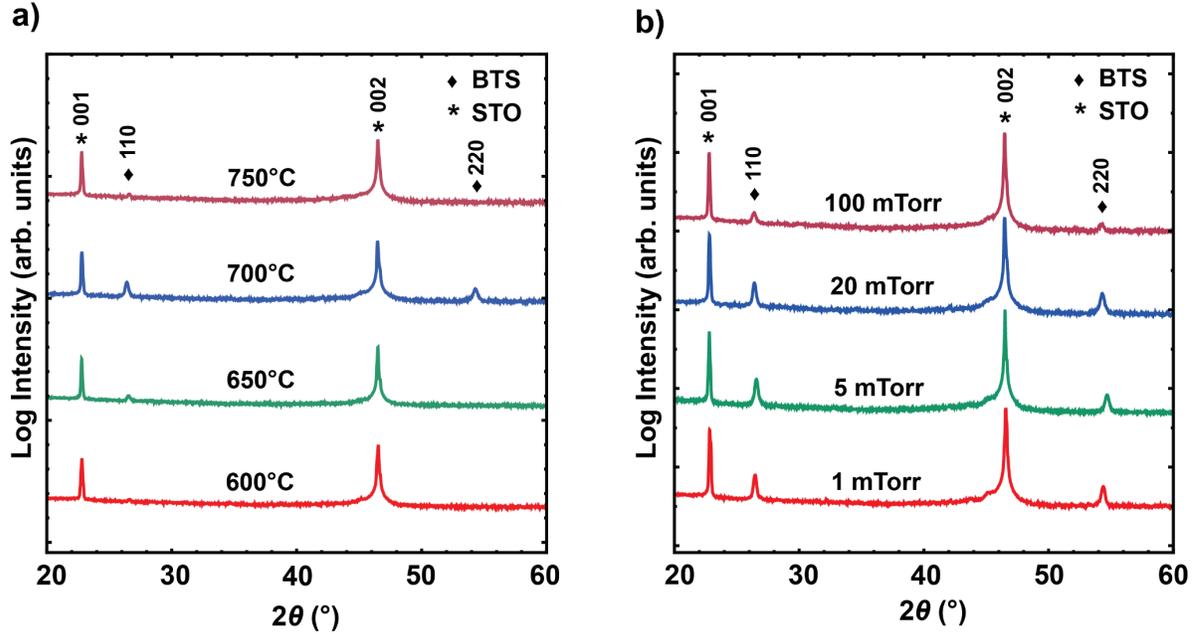

**Figure 2**. Out of plane XRD pattern of BTS films grown on STO substrate at a) substrate temperatures of 600, 650, 700 and 750°C (20 mTorr Ar-$H_2$S partial pressure), and b) 1, 5, 20 and 100 mTorr (substrate temperature of 700 °C).

We further varied the Ar-$H_2$S partial pressure from 1 to 100 mTorr at a fixed temperature of 700°C. Figure 2b illustrates an XRD plots showing the dependence of the Ar-$H_2$S partial pressure on the crystalline quality of the BTS films. The films showed pronounced texture at partial pressures in the range of 1 – 20 mTorr. At high partial pressures of 100 mTorr and above, BTS films were poorly textured, and the 110-reflection peak disappears likely due to the elevated interface roughness caused by the interactions of the reactive $H_2$S and/or ionized Ar. On the other hand, at very low partial pressures, the films were light in color and showed weak texture presumably due to low concentrations of $H_2$S. The rocking curve full-width half maximum (FWHM) decreases at lower pressures with a minimum of 0.92° at 1 mTorr. This suggests that better crystalline quality BTS films are grown at relatively lower partial pressures of H₂S when the interface damage or roughening caused by Ar-H₂S is minimal.

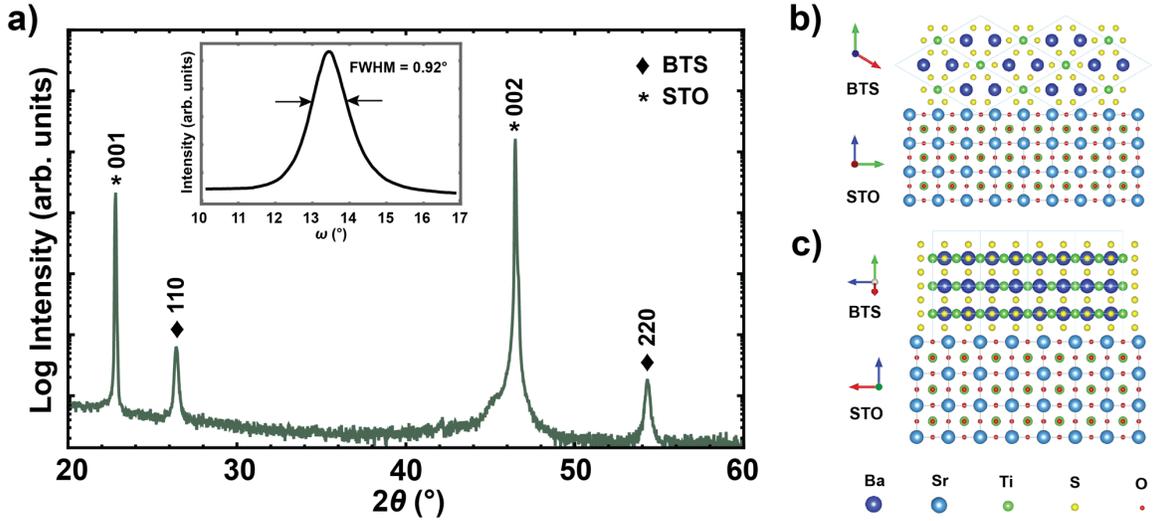

**Figure 3. a)** High resolution 2$\theta$-$\theta$ XRD pattern of a representative BTS film on STO substrate, inset shows the rocking curve of BTS 110 reflection with an FWHM of 0.92°, and schematic showing two different orientations of BTS, b) [001] and c) [1$\bar{1}$0] or [100] along [100] STO.

Figure 3a shows a representative high resolution out-of-plane 2$\theta$-$\theta$ HRXRD scan of a BTS film grown on STO substrate at 700°C and 1 mTorr (optimized growth conditions) where only *hh0* reflections are visible. This indicates that the BTS film is highly oriented along the out-of-plane direction of the STO substrate. Assuming a pseudo-cubic unit cell for BTS with [110] as out of plane direction, the in-plane directions will be and [1$\bar{1}$0] and [001]. In that case, BTS will have a very large lattice mismatch of along STO [100] and STO [110] in-plane direction. Conventional heteroepitaxy requires the film to be strained to match the lattice spacings of the film and the substrate, which is not observed in our BTS thin films. However, energetically favorable interfacial orientations can be attained if the misfit is minimized by a near coincident site lattice matching between the film and the substrate[28]. In many cases, such a near coincident site lattice matching can be accomplished by allowing small in-plane rotations of the film with respect to the substrate. This unconventional near coincident site lattice epitaxy or quasi-epitaxy (we will use this terminology for brevity) is possible when [001] or [100] (or [1$\bar{1}$0]) BTS is grown parallel [100] direction of STO, where resulting in a lattice mismatch of ~1% and 0.1% respectively (lattice mismatch, f =100% x (1- 2$d_{[001]\,BTS}$ /3$d_{[100]\,STO}$)). A schematic of the two possible in-plane orientations of BTS viewed along [100] direction of STO is shown in figure 3b-c. The accessibility to extremely small lattice mismatch can potentially make way for an incommensurate epitaxial growth mechanism. This is perhaps why BTS thin films exhibit the strongest texture on STO substrates, compared to other substrates chosen in this study. The rocking curve of 110 reflection had an FWHM of about 0.92° as shown in the inset of figure 3, indicating the presence of defects and therefore, the crystalline quality of the films need further improvement. Although strongly textured, the peak intensity for the *hh0* reflections were considerably lower and the rocking curve FWHM was about an order of magnitude higher compared to typical similar structured thin films such as perovskite oxides. This indicates that the misorientation caused by extended defects leads to significantly smaller crystalline domains compared to epitaxial oxides, which could be a consequence of film growth in a corrosive $H_2S$ environment along with damage induced by the argon ions in the plasma plume.

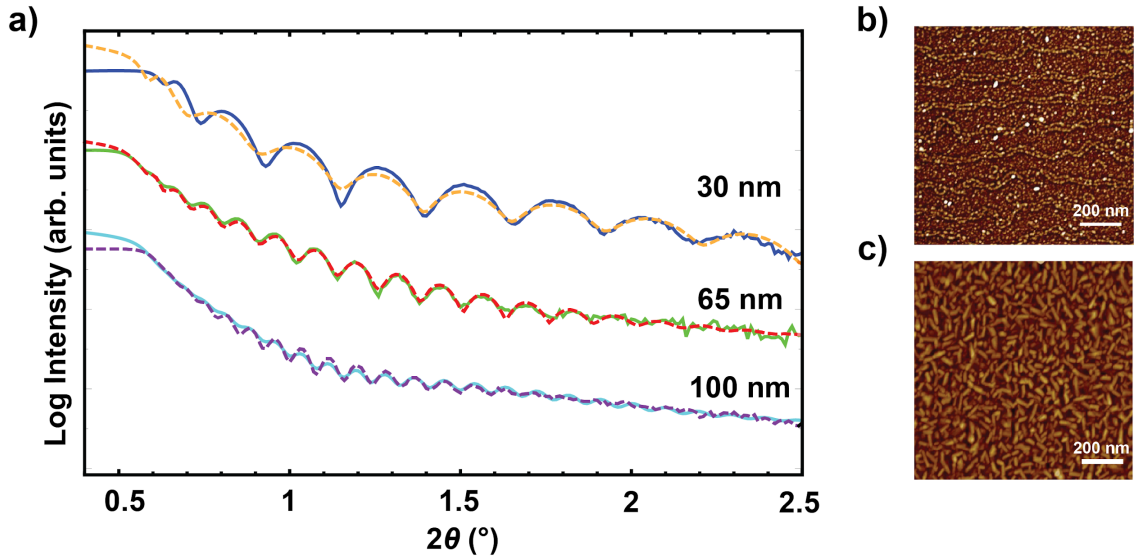

**Figure 4.** (a) XRR patterns showing BTS films grown on STO 001 substrate with varying thicknesses. The corresponding AFM images of (b) 30 nm and (c) 95 nm thick BTS film are also shown.

To analyze the interface and surface roughness of BTS films, we performed X-ray reflectivity (XRR) and atomic force microscopy (AFM) studies on different samples with varying thicknesses. The thicknesses of the films were measured by XRR as shown in Figure 4a. The slow decay of the reflected X-ray intensity and the presence of Kiessig fringes indicate that the films have smooth surfaces and interfaces. However, we observed that the film roughness increased from less than 1 nm for a 30 nm thick film to about 4 nm for a 95 nm BTS film as shown by the AFM topography images shown in figures 4b and c. This relatively high roughness could be attributed to the interface damage caused during the thin film growth in the Ar and $H_2S$ atmosphere. Next, we carried out extensive structural characterization to understand the in-plane crystalline nature and orientation of the BTS films.

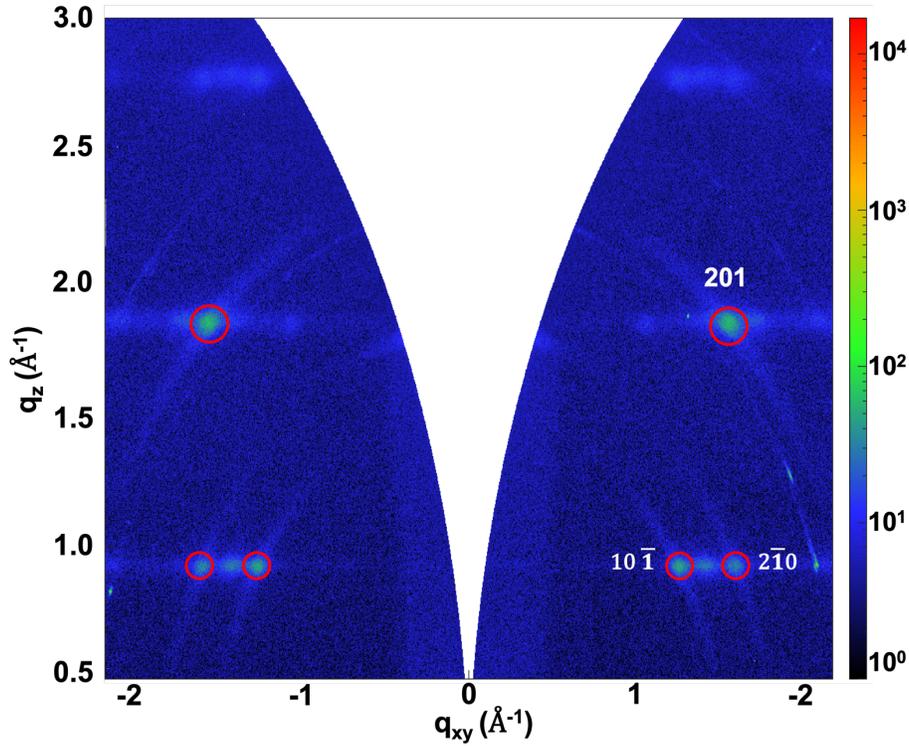

**Figure 5**. GIWAXS pattern of a 100 nm BTS film at an incident angle of 0.5° showing different grain orientations present.

We have further performed off-axis pole figure scans on BTS to investigate the epitaxial relationship of the film and the substrate. However, we did not observe any direct evidence for relationship in the high-resolution thin-film diffraction measurements. If the structural domain sizes are too small, the diffraction peak intensities will be extremely diminished and broad, and hence, it will be difficult to observe any off-axis reflections in typical laboratory scale diffractometers. Therefore, we resorted to grazing incidence wide-angle X-ray scattering (GIWAXS), a conventional X-ray scattering technique, to probe the crystal structure of the BTS thin films[29-31]. The state-of -the -art synchrotron- based GIWAXS technique is usually employed for this purpose, however, due to the high scattering cross-section of perovskites, an easily accessible, laboratory scale GIWAXS with an in-house X-ray source can also be used. The major advantage of the GIWAXS characterization is the 2D area detector which can acquire a simultaneous signal collection in both the in-plane and the out-of-plane directions. Hence, it provides additional information about the orientational order of disordered, polycrystalline films. Figure 5 shows the GIWAXS pattern of a 100 nm BTS film on STO.

An inaccessible region or a "missing wedge" exists in the GIWAXS pattern in Figure 5, owing to the curvature of the Ewald sphere. The diffraction signals from highly ordered crystallites may be hidden in the inaccessible zone. Therefore, the out-of-plane 110 reflection at q=1.86 Å$^{-1}$, which was clearly observed in the HRXRD, was not observed in our GIWAXS studies. We observed multiple discrete spots in the GIWAXS pattern suggesting the presence of grains with different orientations. The discrete diffraction spot located at q=2.4 Å$^{-1}$ with $\chi = 39.4°$ corresponds to the lattice planes with a *d*-spacing value of 2.61 Å and at an inclined angle of about 39.4° from the surface. This peak can be indexed to 201 reflection of BTS as shown in figure 5. Additional reflections at q = 1.84 Å$^{-1}$ with $\chi = 60°$ and q = 1.52 Å$^{-1}$ with $\chi = 52.7°$ were also detected, which correspond to $2\bar{1}0$ and $10\bar{1}$ respectively. Although several grain orientations were observed, <201> reflections appear to be the most intense, which corresponds to an in-plane direction of $[1\bar{1}1]$ BTS. If the proposed incommensurate epitaxy is present, BTS $[1\bar{1}1]$ direction will be parallel to STO [110] in-plane with a small lattice mismatch (1- $2d_{1\text{-}11}/3d_{110}$ = 0.5%). Thus, the presence of a strong <201> reflections could be indicative of an incommensurate relationship.

We also performed a detailed scanning transmission electron microscopic (STEM) study to characterize the interface between the BTS film and STO substrate with atomic resolution and check for the existence of an epitaxial relationship. Figure 6a shows a wide field-of-view high-angle annular dark field (HAADF) image of the ~110 nm thick BTS film on the STO substrate. The BTS films were polycrystalline with a thin (1 – 2 nm) amorphous STO layer formed at the substrate-film interface. This could be attributed due to the formation of defects close to the STO substrate heterointerface due to the high temperature growth in corrosive $H_2S$ environment/damage from Ar ions. Despite the amorphous interfaces, as shown in Figure 6b-c, we observed two typical orientations of BTS were also found to be parallel to [100]-STO. Figure 6b shows an atomically resolved HAADF image of a [001]-oriented BTS grain — that was taken from the region highlighted with a blue square in Figure 6a —, together with a fast Fourier transform (FFT) of the image. The 6-fold symmetry is clearly visible in the FFT pattern. In HAADF images, the intensity of the atomic columns is proportional to the atomic number squared ($Z^2$)[32]. Hence, the heaviest Ba atomic columns appear brightest followed by Sr and Ti. The O and S atomic columns are not visible due to the limitations of the dynamic range of the detector. Figure 6c shows a HAADF image and the corresponding FFT pattern of another BTS grain — that was taken from the region highlighted with a yellow square in Figure 6a. This grain is orientated along the [100] direction of BTS and shows a 2-fold symmetry. Based on the orientation analyses from FFT patterns, we have determined the two epitaxial relationships between the polycrystalline BTS film and the STO substrate as: (110) [001] BTS // (001) [100] STO and (010) [100] BTS // (001) [100] STO.

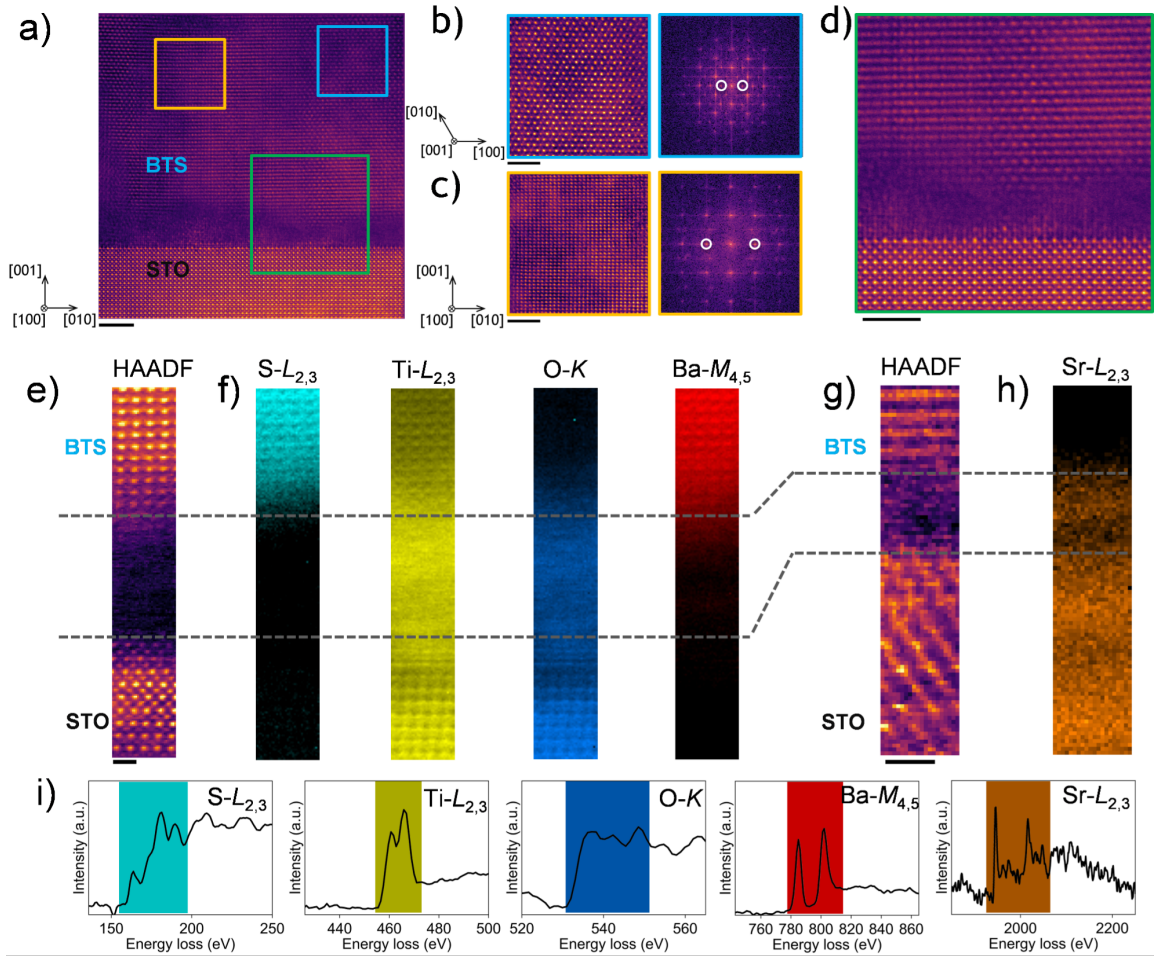

Figure 6. (a) A wide-field-of-view HAADF image showing the polycrystalline BTS film grown on STO substrate. (b-c) Atomic resolution HAADF images along with corresponding FFT patterns showing the two predominant growth orientations of BTS. These images were acquired from the regions highlighted with orange and blue boxes in (a). The reflection spots circled in FFT patterns are corresponding to $(110)/(1\underline{1}0)$ pairs along [001] zone axis in (b), and $(030)/(0\underline{3}0)$ pairs along [100] zone axis in (c). (d) An enlarged HAADF image from the region highlighted with a green box in (a) showing the interface between the STO substrate and polycrystalline BTS. The amorphous region at the interface is clearly visible. (e) HAADF image of the BTS/STO interface that was simultaneously acquired with the EEL spectrum images shown in (f). (f) Elemental mapping using Ti-$L$, O-$K$, S-$L$ and Ba-$M$ edges in the EEL spectrum. (g) HAADF image of the BTS/STO interface that was simultaneously acquired with the EEL spectrum image shown in (h). (h) Sr-$L$ edge map of the BTS/STO interface. (i) The EEL spectra of corresponding elements used to make the elemental maps in (f) and (h) after PCA denoising and back-ground subtraction. The energy range used for signal integration have been highlighted with colors. Scale bars represent 2 nm in (a-d) and 0.5 nm in (e) and (g).

Our STEM analysis also indicates the existence of a thin amorphous layer at the heterointerface between BTS and STO. Figure 6d shows an atomic resolution HAADF image of the BTS/STO interface — that was taken from the region highlighted with a green rectangle in Figure 6a. Even in those areas where BTS with thin amorphous STO layer, the orientation relationships between BTS and STO still maintain as the same stated above. We suppose the amorphous layer is formed due to the corrosive damage of STO substrate caused by $H_2S$ during BTS growth at high temperature and/or ionized argon ions. Thus, we show that the proposed quasi-epitaxial relationship exists despite the presence of an imperfect substrate-film interface. Additionally, we have also identified local regions of BTS films where the (110) [001] BTS // (001) [100] STO epitaxial relationship was observed without an amorphous region. This suggests that if the Ar-$H_2S$ induced damage during the deposition can be avoided, heteroepitaxial growth of BTS can be achieved on STO substrates with a sharp interface and *c*-axis oriented in the in-plane direction. This is desirable for exploiting the large optical anisotropy of BTS for polarization dependent infrared optics and optoelectronics.

We have further performed electron energy loss spectroscopy (EELS) to map the distribution of elements across the heterointerface. Figure 6e shows a HAADF image of the interface where EELS data was simultaneously acquired for mapping the distribution of Ti, O, S, and Ba. Figure 6(f) shows the elemental maps of Ti-*L*, O-*K*, S-*L*, and Ba-*M* edges. As the Sr-*L* edge has a much higher absorption onset at ~1940 eV than the other elements (see Figure 6i), we acquired the Sr-*L* edge spectrum image from a region of the interface adjacent to where the spectrum images of the other elements shown in 6(f) were obtained. Figure 6(g-h) shows the HAADF image and the simultaneously acquired Sr-*L* edge map. From the Sr, Ti, S and O maps and the HAADF images, it is clear that the amorphous layer is a damaged STO substrate region and the BTS layer is crystalline. From the Sr-*L* edge and Ba-*M* edge elemental maps, we do not observe large intermixing of the two elements across the heterointerface. The oxygen content in the bulk of the film was below the measurable limit. In Figure 6i, we are showing the extracted core-edges after background-subtraction of the five elements together with the energy ranges over which the respective signals were integrated to obtain the elemental maps.

The limited availability of structurally similar and lattice-matched substrates and non-corrosive chalcogen precursor are the major challenges in the epitaxial thin film growth of BTS. New strategies need to be devised to improve the substrate- film interface and thus obtain high-quality epitaxial films with large grain growth. Optical and optoelectronic characterization of BTS thin films is a part of our ongoing work. This study opens up opportunity to investigate the fundamental properties and realize exciting optoelectronic and photonic devices in infrared chalcogenide perovskite thin films and heterostructures.

**Conclusion**

Quasi-epitaxial BTS thin films has been demonstrated by pulsed laser deposition. The growth parameters such as the substrate temperature and $H_2S$ partial pressure were varied and their effects on the thin film structure were examined, as measured by X-ray diffraction, to understand the effect of the process parameters on the microstructure of the BTS thin films. X-Ray diffraction shows strong texture in the out-of-plane direction, whereas no evident in-plane relationship between the film and the substrate is observed by laboratory scale high resolution diffractometer. GIWAXS and STEM studies revealed the presence of two epitaxial relationships between the film and substrate, despite the existence of a thin amorphous layer at the interface - (110) [001] BTS // (001) [100] STO and (010) [100] BTS // (001) [100] STO. This work emphasizes on the challenges in the thin film growth of high-quality epitaxial chalcogenide perovskite, such as lack of easily available epitaxial substrates, cation-chalcogen vapor pressure mismatch and the use of corrosive background gas. It is extremely important to improve the film-substrate interface during the growth to achieve high-quality epitaxial films and is a part of our ongoing studies.

## Experimental Section

**Thin film deposition**

The BTS thin films were grown by PLD using a 248 nm KrF excimer laser in a hydrogen sulfide compatible vacuum chamber specifically designed for the growth of chalcogenide perovskites. Single crystal perovskite oxide (Crystec GmbH) substrates such as STO, LAO, LSAT and sapphire were pretreated by annealing at 1000°C for 3 h in 100 sccm $O_2$ and subsequently cleaned in acetone and IPA prior to deposition. The chamber was evacuated to a base pressure of ~ $10^{-8}$ mbar and then backfilled with high purity argon gas. The substrate was heated up to the growth temperature in an argon partial pressure of 10 mTorr. A dense poly crystalline 1-inch BTS pellet synthesized by high pressure torsion process was used as the target, which was preablated before growth. Argon – $H_2S$ (4.75%) gas mixture was introduced into the chamber right before the growth to achieve growth pressures in the range of 1-20 mTorr. The fluence was fixed at 1.0 J/cm$^2$ and the target-substrate distance used was 75 mm. The films were cooled postgrowth at a rate of 5°C/min at an Ar-H$_2$S partial pressure of 200 mTorr.

**X-ray diffraction and reflectivity**

The high resolution out of plane XRD scans were carried out on a Bruker D8 Advance diffractometer using a Ge (004) two bounce monochromator with Cu Kα1($\lambda$ = 1.5406 Å) radiation at room temperature. XRR measurements were done on the same diffractometer in a parallel beam geometry using a Göbel (parabolic) mirror set up.

**Atomic Force Microscopy**

AFM was performed on Bruker Multimode 8 atomic force microscope in peak force tapping mode with a ScanAsyst tip geometry to obtain the surface morphology and roughness.

**Grazing incidence wide-angle X-ray scattering**

GIWAXS experiments were conducted in vacuum conditions and at room temperature using a Xenocs Xeuss 3.0 laboratory instrument with an incident wavelength of 1.5406 Å. An incident angle of 0.5° was used for GIWAXS. X-ray scattering data was recorded on a Pilatus 300k area detector at a sample-detector distance of 72 mm.

**Electron microscopy Sample preparation**

STEM sample preparation was performed using a Thermo Scientific Helios G4 PFIB UXe Dual Beam equipped with an EasyLift manipulator. Standard in situ lift-out technique was used to prepare the TEM lamella. The lift-out region was first coated with the thin e-beam deposited W and followed by a thicker 2μm ion-beam deposited coating to avoid ion-beam damage to the BTS film. The sample was milled and thinned down to ≈100nm using a 30kV beam and final polishing to remove residual surface beam damage was carried out using a 3kV ion beam.

**Scanning transmission electron microscopy**

Atomic-resolution STEM imaging and EEL spectroscopy was performed using the aberration-corrected Nion UltraSTEM™ 100 at CNMS, ORNL operated at 100 kV with a convergence angle of 30 mrad. HAADF-STEM images were acquired on the annular dark field detector with a collection angle of 80-200 mrad. Core-loss EELS spectra were collected using a Gatan Enfina EEL spectrometer. A collection semi-angle of 48 mrad and an energy dispersion of 1 eV per channel were set for EELS acquisition. To improve the signal-noise ratio, we have applied principal component analysis (PCA) to EEL spectrum images. A power law was used to model the background signal prior to each core-loss signal.

**Acknowledgements**

This work was supported in part by the Army Research Office under Award No. W911NF-19-1-0137, an ARO MURI program with award no. W911NF-21-1-0327, the National Science Foundation of the United States under grant numbers DMR-2122070 and DMR-2122071, and an Air Force Office of Scientific Research grant no. FA9550-22-1-0117. STEM characterization was conducted at the Center for Nanophase Materials Sciences at Oak Ridge National Laboratory (ORNL), which is a Department of Energy (DOE) Office of Science User Facility, through a user project (G.D.R. and R.M.). The work of HPT processing at Oregon State University was supported by the National Science Foundation of the United States under Grant No. DMR-1810343. The authors gratefully acknowledge the use of facilities at the Core Center for Excellence in Nano Imaging at University of Southern California for the results reported in this manuscript.

**Conflict of Interest**

The authors declare no conflict of interest.

**Data Availability**

The data are available from the corresponding authors of the article on reasonable request.